\pdfoutput=1
\documentclass[%
 reprint,
 amsmath,amssymb,
 aip,
 rsi,
]{revtex4-2}

\usepackage[T1]{fontenc}
\usepackage{graphicx}
\usepackage{dcolumn}
\usepackage{bm}
\usepackage{hyperref}
\usepackage{xcolor}
\usepackage[per-mode=symbol]{siunitx}
\usepackage{booktabs}

\newcommand{\labscript}{{\texttt{labscript}} }

\begin{document}

\preprint{APS/123-QED}

\title{Experimental timing and control using microcontrollers
}

\author{Philip T. Starkey}
\noaffiliation
\author{Carter Turnbaugh}%
\affiliation{%
 Department of Physics, MIT-Harvard Center for Ultracold Atoms, and Research Laboratory of Electronics, MIT, Cambridge, MA 02139 USA
}%

\author{Patrick Miller}
\affiliation{%
 Department of Electrical and Computer Engineering, University of Maryland, College Park, MD 20742 USA
}%
\author{Kermit-James LeBlanc}
\affiliation{%
 DEVCOM Army Research Laboratory, Adelphi, MD 20783 USA
}%
\author{David H. Meyer}
\email{david.h.meyer3.civ@army.mil}
\affiliation{%
 DEVCOM Army Research Laboratory, Adelphi, MD 20783 USA
}

\date{\today}

\begin{abstract}
Modern physics experiments rely on precise timing provided by programmable digital pulse generators.
In many experimental control systems, this role is filled by custom devices built on Field Programmable Gate Arrays (FPGAs). 
While highly flexible and performant, these systems can be difficult to scale to very large systems due to cost and complexity.
Recent advances in microcontroller systems allows these much simpler systems to fill the role of digital pulse generators.
Here we demonstrate one such alternative based on the Raspberry Pi Pico microcontroller which allows for timing resolution down to \qty{7.5}{\ns} with a minimum pulse width of of \qty{37.5}{\ns}.
\end{abstract}

\maketitle


\section{Introduction}

Experiment control in physics relies on precise, repeatable, reconfigurable timing provided by digital pulse sequences.
Arbitrary pulse sequences can be used to implement hardware-timing for an experiment by providing digital pulses derived from a stable system clock to produce triggering events at precisely-timed intervals.
These trigger signals are then used to control inputs and outputs of other devices that control system behavior.
Examples common in atomic, molecular, and optical (AMO) physics experiments include: 
toggling of radio-frequency switches,
periodic clock sequences that trigger sampling in analog-to-digital or digital-to-analog converters,
or gating of single photon counters.
As experiments grow in complexity and size, particularly in quantum information science, the requirements on the digital timing and control signals must also scale and improve in performance.

Field-Programmable Gate Arrays (FPGAs) are often used to fulfill this role \cite{bertoldi_control_2020,perego_scalable_2018,donnellan_scalable_2019,keshet_distributed_2013,lee_new_2013,kulik_driver_2018,kasprowicz_artiq_2020,sitaram_programmable_2021}.
Their reconfigurable nature allows for a high degree of flexiblity, making them suitable to handle many disparate tasks simultaneously.
FPGAs often have significant capacity in Input/Output (IO) channels and can have very fast system clock speeds allowing for fast operation and high timing resolution.
Finally, it is possible to implement mid-experiment branching logic within the hardware, allowing for highly dynamic experiment control sequences that operate with the speed and precision of hardware timing.

However, there are costs associated with using an FPGA.
Beyond outright expense and power usage, both of which can be high for the highest performing models, FPGAs are more complex and time-consuming to program.
They require specialized hardware device languages and, frequently, proprietary software development kits with associated licensing terms that can change and/or limit development.
Moreover, developed programs can be difficult to transfer between hardware implementations, which can make replacing discontinued FPGA chips challenging.

Ultimately, the processes of digital pulse generation for experiment timing and device triggering often does not require the full power of an FPGA device.
In many situations, a capable microcontroller unit (MCU) is sufficient \cite{patel_experimental_2020,hosak_arbitrary_2018,gaskell_open-source_2009,malek_embedded_2019}.
Considering their reduced cost, power consumption, and relative simplicity in programming (using standard programming languages such as C, C\texttt{++}, or even Python) a microcontroller can be an effective, scalable alternative for arbitrary digital pulse generation.

The RP2040\cite{rp2040datasheet} that powers the Raspberry Pi Pico board\cite{picodatasheet} is a new and highly capable microcontroller that contains many features that make it well suited to digital pulse generation requirements in experiment control, such as parallel, high speed processing cores and dedicated hardware for memory management and digital output control \cite{huegler_agile_2023}\footnote{References to commercial devices do not constitute an endorsement by the US Government or the Army Research Laboratory.}.
Here we will describe two custom firmwares for the Raspberry Pi Pico microcontroller board: the Prawnblaster and the PrawnDO \footnote{The PrawnBlaster is named after the Australian \$5 note, which is colloquially known as a ``Prawn''.}.
Each implements a distinct method of digital pulse generation often required of experiment control: the Prawnblaster produces pseudoclock pulse sequences with programmable periods and number of pulses, and the PrawnDO implements arbitrary pulse generation.
Used together, they provide full programmable digital pulse generation capability that is highly scalable.
Though intended to be agnostic to experiment control implementation, both boards are designed for simple integration into the labscript suite which employs pseudoclock sequences for experimental timing as well as general purpose digital outputs \cite{starkey_scripted_2013}.

\section{System Overview}

Our system is broken into two distinct methods of digital pulse generation in order to best optimize the programming of arbitrary pulse sequences required in a typical experiment.
Each method is implemented by a distinct firmware for the Pico board, with the primary distinction between them being how the pulse sequence is quantified or stored in memory.
Each board is designed to be readily interconnected to provide more outputs of each type.
This framework leverages the minimal cost of the Pico board, where tens of microcontrollers can be purchased for the same cost as a single FPGA.

\begin{figure}[t]
    \centering
    \includegraphics{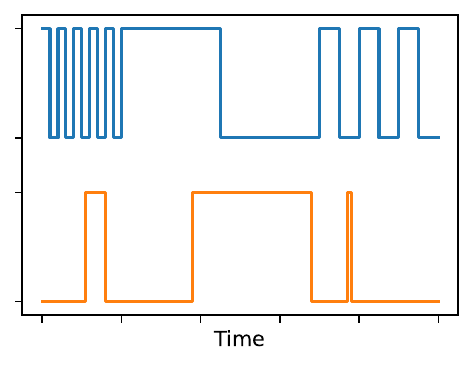}
    \caption{Example digital waveforms.
    Upper blue trace is a pseudoclock waveform (described in \ref{subsec:pseudoclocks}). It can be specified with three commands: five short pulses, a single long pulse, and three intermediate length pulses.
    Lower orange trace is a typical arbitrary pulse waveform (described in \ref{subsec:arbpulses}). It requires seven instructions, one for each change of state.}
    \label{fig:pulsetypes}
\end{figure}

\subsection{Pseudoclocks}
\label{subsec:pseudoclocks}

The first pulse generation method is the pseudoclock \cite{starkey_scripted_2013}.
A pseudoclock is much like a typical clock signal: it uses digital pulses to implement a waveform with a 50/50 duty cycle and a fixed period.
Unlike a typical clock, the pseudoclock produces sequential pulse sequences with a fixed number of pulses and variable period.
In other words, sequential pulse sequences are specified by the number of pulses of a fixed period.
The example pseudoclock waveform shown as the upper trace of Figure \ref{fig:pulsetypes} can be specified using only three commands:
five pulses with a short period,
a single pulse with a long period,
followed by three pulses of an intermediate period.

Quantifying timing waveforms in this way has a number of advantages.
Many devices, such as analog-to-digital or digital-to-analog converters, require large numbers of repetitive triggers to control the sampling rates when producing or capturing dynamic signals.
A pseudoclock allows for a single instruction to command such sampling clocks, and the variable nature allows for a single digital output to alter the sampling rate as needed within a single experiment.
Allowing variable sampling rates also avoids unneeded analog-to-digital and digital-to-analog conversions, reducing the number of instructions and/or measurements that devices need to handle.

This method is also flexible enough to produce single trigger pulses by specifying a sequence of one pulse. However, as this method produces pulses with a 50/50 duty cycle, it cannot produce an arbitrary pulse.

\subsection{Arbitrary Digital Pulses}
\label{subsec:arbpulses}

The second pulse generation method is focused on producing arbitrary pulses for single trigger events (such as an oscilloscope capture), gating, or switching controls.
These types of timing events often require fully arbitrary pulses, with full control over both the high and low times. 
An example of such a sequence is shown as the lower trace of Figure \ref{fig:pulsetypes}.
There are multiple equivalent methods of parameterizing this sequence.
We use run-length encoding, where each instruction includes the state of the output (high/low) and how long to maintain that state.
While more flexible than a pseudoclock, even the fairly limited sequence of pulses in Fig.~\ref{fig:pulsetypes} requires seven instructions, one for each rising/falling edge.

The most important advantage of using both types of pulse generation in a single system is that we can separate devices that require different types of digital triggers (pseudoclock versus arbitrary pulses) and use optimal instruction sets for both situations.
By moving devices that require clock-like timing triggers to a pseudoclock where they can defined with minimal instructions, the more instruction-heavy arbitrary pulses are often limited to devices that only need a handful of pulses.
This optimization is critical when using microcontrollers that are often limited to low speed serial communication or have limited local memory.

\begin{figure}[tb]
    \centering
    \includegraphics{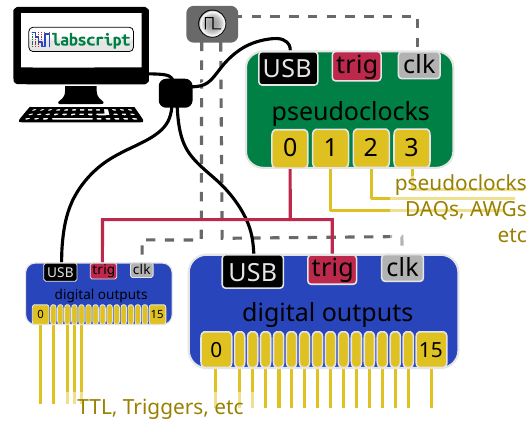}
    \caption{Digital pulse generation system overview.
    System consists of inter-linked microcontrollers of two types: pseudoclock generating Prawnblasters and arbitrary digital output generating PrawnDOs.
    All boards are controlled from a parent host-controller that programs all devices.}
    \label{fig:system}
\end{figure}

Figure \ref{fig:system} shows an overview of the physical wiring on an example digital pulse generation system.
A control computer (in our case using \texttt{labscript} as the experimental control software) uses Universal Serial Bus (USB) connections to each Pico board.
A parent Prawnblaster board is then used to provide triggering to PrawnDO boards that can provide the required ancillary digital outputs.
In this example configuration, multiple PrawnDO boards are triggered in parallel from the same trigger pseudoclock output, increasing the total number of digital outputs available.
Ultimate choice of topology is dictated by experimental requirements.
Finally, in order to ensure consistent timing between multiple boards, a common system reference clock is provided to each.

\section{Hardware}

The Raspberry Pi Pico microcontroller board\cite{picodatasheet}, which uses the RP2040 microcontroller \cite{rp2040datasheet}, provides the hardware backbone to our digital pulse generation system.
Despite the low-cost, this microcontroller contains a number of key features that allow us to produce arbitrary digital pulses with timing resolution at the \qty{10}{\ns} level.

\begin{itemize}
    \item A system clock of up to \qty{133}{\MHz}, allowing for timing resolution of pulse edges down to \qty{7.5}{\ns}.
    \item Dual-core central processing unit which allows dedicating one processor to handle interfacing with the controlling computer and the other to control pulse generation resources.
    \item Dedicated Programmable IO (PIO) cores for direct real-time control of General-Purpose IO (GPIO) outputs independent of the central processor. Operation is programmed in a specialized subset of assembly, allowing for precise timing of instructions relative to the system clock frequency.
    \item \qty{264}{\kilo\byte} of internal Static Random-Access Memory (SRAM) providing ample local storage of the firmware along with 30,000 pulse instructions, reducing the need for external memory.
    \item Dedicated Direct Memory Access (DMA) controller that handles the required, repetitive, fast memory transfers between system memory and PIO cores without requiring the central processor.
    \item USB controller and physical layer capable of operating as a USB Full Speed device, with \qty{12}{\mega\bit\per\second} bandwidth. This enables writing instructions much faster than with a USB-to-UART (Universal Asynchronous Receiver-Transmitter) adapter.
\end{itemize}

The Prawnblaster and PrawnDO firmwares use these resources to implement digital pulse generation with timing resolution of a single system clock cycle, and a minimum pulse period of five clock cycles.
For a \qty{100}{\MHz} system clock, this means timing resolution of \qty{10}{\ns} with a minimum pulse period of \qty{50}{\ns}.

\begin{figure}[tb]
    \centering
    \includegraphics[width=0.8\linewidth]{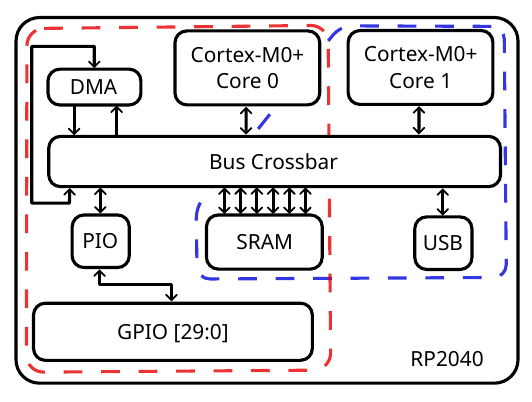}
    \caption{Simplified diagram of the RP2040 microcontroller sub-systems the digital pulse generation relies upon.
    The blue outlined systems are used for the serial interfacing with the host controller. The red outlined systems are used for digital pulse generation using the PIO core state machine(s).}
    \label{fig:RP2040}
\end{figure}

Figure \ref{fig:RP2040} shows a simplified block diagram of these RP2040 sub-components.
The blue outline denotes the serial communication resources (USB controller, SRAM for instruction storage, and one of the processing cores), while the red outline denotes the digital pulse generation resources (PIO core, GPIO outputs, DMA controller that moves stored instructions from SRAM to the PIO).
The bus crossbar provides the interconnects that route addresses and data between the various sub-components of the microcontroller.
It has four upstream ports (ie devices that can generate addresses: the processing cores and the DMA read/write channels) and ten downstream ports that route to varous sub-components (such as the PIO cores, SRAM modules, and the USB controller).
The crossbar supports up to four simultaneous bus transfers per clock cycle, with each data path being 32-bits wide.

While specifications for the Prawnblaster and PrawnDO firmwares will be discussed in the next section, here we will discuss the basic operation and specifications of the RP2040 sub-components that influence digital pulse generation performance.

The RP2040 includes \qty{264}{\kilo\byte} of SRAM that is used to store the compiled bytecode as well as any user defined instructions.
Though the Pico board contains another \qty{2}{\mega\byte} of flash memory, our firmwares only use the MCU SRAM.
This simplifies memory management while still allowing for thousands of instructions.

The PIO cores are specifically designed to handle IO tasks that require precise timing and determinism, such as Serial Programming Interface (SPI) or Inter-Integrated Circuit (I2C) protocols.
The RP2040 has two PIO cores that each have direct access to the GPIO and bus crossbar.
Furthermore, each core has four independent state machines, where each state machine can independently process sequential programs that read/write the GPIO or perform data transfers.
These programs are written in a subset of assembly language (ASM) that contains only nine instructions, where each ASM instruction takes a fixed number of clock cycles to perform, allowing for a high degree of granular control and timing precision.
Each state machine also provides shift and scratch registers as well as First-In/First-Out (FIFO) buffers for locally handling data.
The core functionality of the Prawnblaster and PrawnDO is contained in the optimized PIO programs.
The rest of the firmware provides interfaces to the controlling host computer and to start/stop the PIO state machines.

Finally, the RP2040 DMA controller allows for off-loading of repetitive memory transfers from the processor, allowing for much higher speed memory transfers.
It has the ability to read and write up to 32-bits per clock cycle,
and can employ up to 12 independent control channels to handle different tasks.
We use the DMA controller to feed instructions from the SRAM into the PIO's FIFO buffer.
As the PIO consumes instructions from the FIFO,
the DMA automatically replenishes it, 32-bits at a time, until execution is complete.

\section{Firmware}

The Prawnblaster and PrawnDO firmwares are readily divided into two functionalities, each being handled by one of the independent processing cores: interfacing with the host controller computer and managing the PIO state machine.
The source code is available on github for both firmwares, along with pre-compiled binaries \cite{prawnblaster,prawnDO}.

The primary processor core handles host interfacing (via a serial interface) and sending instructions to the secondary processor to start and stop PIO state machine execution.
Communication between cores uses two mutex variables.
One instructs the other processor to start or stop the state machine execution,
the other mutex contains the operational status of the state machine (ie idle, running, aborting, etc).

The host controller interface uses a serial over USB Communications Device Class (CDC) interface to send string commands and responses between the board and the controller.
There are commands for managing the instructions in memory, manually changing output states, status reporting, IO configuration, and starting/stopping execution of the instructions in memory.
Most commands are sent as human readable ASCII-strings for ease of debugging, but instruction writing can also be done via a bulk binary transfer.
As the RP2040 uses USB transfers, software checksums and flow control are not needed, allowing the write speed (using standard serial over USB CDC drivers) to reach approximately \qty{30}{\kilo\bit\per\second} for a series of standard ASCII commands and \qty{650}{\kilo\bit\per\second} for bulk binary transfers.
As a result, writing the full memory in bulk binary mode only requires \qty{270}{\milli\second}.
To reduce overhead from the Pico SDK stdio library, we directly use the TinyUSB library.

The PIO state machine controlling processor configures and manages the PIO that outputs the digital pulses as well as the DMA transfers that move the instructions from memory to the PIO core.
It also monitors execution progress and communicates state to the interfacing processor.
Finally, it listens for abort commands mid-execution so that operation can be interrupted if needed.

A typical sequence of events when using the Prawnblaster or PrawnDO to output digital pulses is as follows:

\begin{itemize}
    \item Send a list of instructions to the device, which are processed and saved in SRAM.
    \item Send the start command. PIO state machine begins execution either immediately (a software start) or after receiving an external trigger (hardware start).
    \item The PIO state machine pulls instructions from the FIFO buffer being populated by automated DMA transfers. Outputs are updated as instructed.
    \item If a wait instruction is encountered, the state machine monitors the trigger input GPIO channel. When that channel transitions to high, execution is resumed.
    \item When the stop instruction is encountered by the state machine, execution ends, the state machine is idled, and successful completion of the program is posted to the device status.
    \item Interfacing computer polls the device status until execution complete is reported. A new set of instructions can now be sent or the same program re-run.
\end{itemize}

As described above, the primary difference between the Prawnblaster and PrawnDO firmwares is how the instructions are defined and processed. 
These differences are discussed in the next two sub-sections.

\begin{figure*}[tb]
    \centering
    \includegraphics[width=0.9\linewidth]{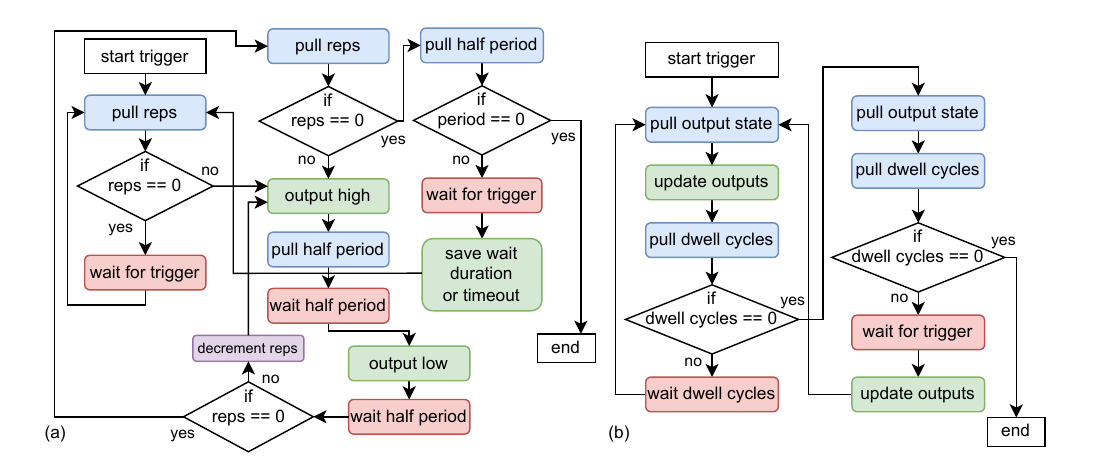}
    \caption{Flow diagram of the PIO cores for the (a) Prawnblaster and (b) Prawn-DO.
    Blue blocks represent data transfers via the DMA controller, red blocks represent dwell events to enforce desired timing, green blocks represent outputs activity, purple blocks represent local modifications of variables, and white diamonds represent branching logic.
    }
    \label{fig:pioflow}
\end{figure*}

\subsection{Pseudoclock Generation - Prawnblaster}

The Prawnblaster is capable of generating up to four independent pseudoclock outputs.
Each output is controlled via one of the four state machines of one of the PIO cores.
As such, the timing of each output is synchronized.
A single instruction for the pseudoclock is made up of two unsigned 32-bit integers.
The first is the half-period, defined as the number of clock cycles to remain high and low for a single period (ie twice the half-period is the full period of the pseudoclock frequency).
The second is the repetitions, defined as the number of pulses to produce before starting the next instruction.
An instruction with repetitions set to 0 indicates a wait or stop instruction, as described below.
The Prawnblaster can hold up to 30000 instructions, evenly distributed between the active pseudoclock outputs.

Valid half-periods are between 5 and $2^{32}-1$, the minimum defined by the minimum time required by the PIO to process the next instruction and the maximum set by the max 32-bit integer (for a \qty{100}{\MHz} system clock, this corresponds to $\qty{\sim42.9}{\s}$).
A half-period of 0 is also valid when repetitions is 0, indicating a stop instruction which ends execution of the program.
Internally, the half period is saved as the programmed value minus five, which accounts for the minimum required number of ASM instructions between output state changes.

There are two types of waits that the Prawnblaster recognizes, a standard wait which includes a timeout, and an indefinite wait which does not.
A standard wait instruction has repetitions set to 0 and the half-period is interpreted as the length of the timeout in number of clock cycles.
For a standard wait, the prawnblaster will wait for a trigger to resume program execution until the timeout has elapsed.
If the timeout elapses without a trigger detected, execution resumes.
The Prawnblaster will save the length of the wait to memory, allowing the user to access it via serial commands after execution has finished.
If two sequential wait instructions are encountered, this indicates an indefinite wait.
Once the timeout of the first wait elapses, a second wait that can only be resumed by an external trigger is begun.

A flow diagram of the PIO state machine code that implements the above behavior is shown in Figure \ref{fig:pioflow}(a).
This code is designed to have the minimum number of ASM instructions in the tightest loop that generates the pseudoclock pulses (center column of the flow diagram), reducing the minimum pulse width to five clock cycles.
As a result, the triggering times and minimum durations surrounding waits are longer.
Waits cannot be shorter than four clock cycles,
and a trigger must be at least four cycles long for a standard wait, and 12 cycles long for an indefinite wait.

\subsection{Arbitrary Digital Pulse Generation - PrawnDO}

\begin{figure*}[t]
    \centering
    \includegraphics[width=0.9\linewidth]{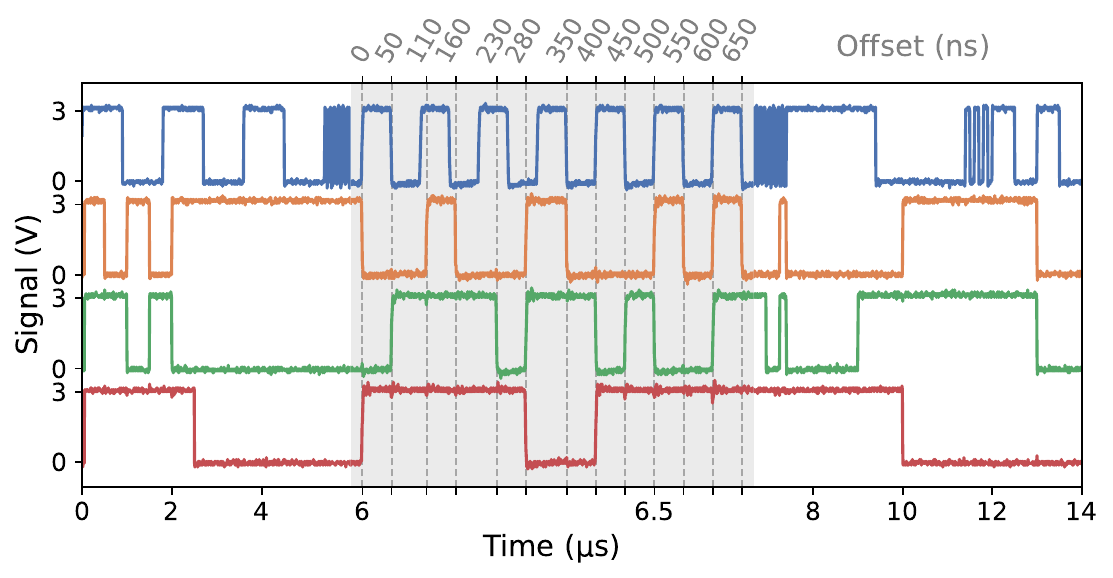}
    \caption{Example digital pulses from a combined pulse generation system consisting of one Prawnblaster and one PrawnDO.
    The top blue trace is a Prawnblaster pseudoclock,
    with the three lower traces being three digital outputs from the PrawnDO.
    The central gray region is a zoomed region of the larger traces, with the gray-dashed vertical lines marking the edges of each state update of the PrawnDO.
    The minimum pulse widths (\qty{50}{\nano\second}) and minimum timing resolution (\qty{10}{\nano\second}) are evident.
    }\label{fig:measTraces}
\end{figure*}

The PrawnDO is designed assuming it will always use an external trigger, such as a Prawnblaster pseudoclock output, to start execution.
As such, its firmware is simpler than that of the Prawnblaster since much of the wait complexity can be reduced by only supporting indefinite waits.
It uses a single state machine of a PIO to control 16 outputs simultaneously.
A single instruction consists of an unsigned 16-bit integer and an unsigned 32-bit integer.
The first is the output states, where the $i$th bit of the integer sets the output state of the $i$th output.
The second 32-bit integer is the number of clock cycles to hold the output state.
For a standard instruction, this must be at least 5 and no more than $2^{32}-1$, same as the Prawnblaster half-period.
If the clock cycles is set to 0, this indicates an indefinite wait: execution will only resume when an external trigger is detected.
If two instructions in a row have clock cycles of 0, this indicates a stop instruction which ends execution of the state machine.

Figure \ref{fig:pioflow}(b) shows the flow diagram of the PIO state machine code for the PrawnDO.
It has the same five clock cycle minimum instruction time as the Prawnblaster, due to the finite number of ASM instructions required to load an instruction and change output.
Like the Prawnblaster, the internally-saved dwell time must also account for this finite number of ASM instructions.

The PrawnDO can hold up to 30000 instructions.
However, note that any change in state of any of the 16 outputs requires an instruction.
If state changes do not align in time, meaning they cannot be programmed as single instructions, even fairly simple pulse programs could lead to very large instruction sets.
As a worst-case example, programming three outputs, each producing a \qty{10}{\MHz} square-wave with a different phase offset such that state changes did not align, would only produce output for \qty{1}{\ms} using the full instruction memory.

Of particular note, the PrawnDO is completely internally timed.
In this context, that means it only requires a (optional) start trigger and re-triggers for waits.
This is in contrast to how similar devices are programmed within the \texttt{labscript} system, where each change would require an external trigger.
This paradigm is generally preferred as it ensures all outputs across the experiment have a consistent time source (the primary pseudoclock).
It also limits the amount of data the device needs to store, as all timing information is handled via external triggers with the parent pseudoclock.
Here we chose to use internal timing, relying on a common external clock to maintain timing stability between the pseudoclock and the PrawnDO.
The limitation this circumvents is the design of \texttt{labscript} which uses only rising \emph{or} falling edges for triggering, meaning that each rising and falling edge of the PrawnDO output would have to be triggered by a rising pseudoclock edge, which has a minimum separation of ten clock cycles in the Prawnblaster.
By relying on a common clock to synchronize the PrawnDO to the Prawnblaster instead, we can leverage the full speed of the PIO to produce pulses at the same minimum of five clock cycles.
Under a typical configuration where the default \qty{100}{\MHz} system clock of the Prawnblaster is used to clock the PrawnDO, we can produce pulses with a minimum duration of \qty{50}{\ns}.

\section{Example Pulse Sequence}

Figure \ref{fig:measTraces} shows an example pulse sequence with a pseudoclock and arbitrary digital pulses from a Prawnblaster and PrawnDO, both using their default \qty{100}{\mega\hertz} internal clocks.
The entire sequence spans \qty{14}{\micro\second}.
The top blue trace is the pseudoclock output and represents five total instructions.
The orange, green, and red traces are outputs 0, 1, and 2 of a PrawnDO that was triggered from another output of the PrawnBlaster.
Its program has 26 instructions; one for each state change on any output.
The precise instructions used are listed in App.~\ref{app:programs}.

The central gray region of Figure \ref{fig:measTraces} shows a zoomed region of the full timetrace covering 6 to \qty{6.8}{\micro\second}.
In this window the Prawnblaster pseudoclock has its minimum period of \qty{100}{\nano\second}.
The top axis shows the times of each update of the PrawnDO relative to \qty{6}{\micro\second}.
This sequence demonstrates the \qty{50}{\nano\second} minimum time between updates as well as the \qty{10}{\nano\second} resolution (e.g. the third instruction is \qty{60}{\nano\second} after the second).

\section{Discussion}

While the split-board design of our system has allowed for a simpler, more performant design, it also allows for simple scaling of the number of outputs as required by an experiment.
Both the Prawnblaster and the PrawnDO are designed to be triggered from other Prawnblasters, meaning that a branching web of triggering outputs can create large numbers of pseudoclocks and arbitrary digital outputs.

However, the distributed design does lead to complexities when connecting boards that would often be mitigated by features found in more expensive devices that the microcontroller boards lack.
In particular, maintaining phase synchronous timing between boards requires distributing a common clock reference to each board.
Also, each board requires a separate USB connection to the host computer.
Finally, all inputs and outputs of the Pico boards are LVCMOS \qty{3.3}{\volt}.
As a result, interfacing with TTL logic or highly capacitive loads like long coaxial cables requires external unity-gain buffers to provide logic level interoperability and higher sourcing currents.
We have designed simple breakout boards that provide these buffers as well as SMA connections.
A more detailed description is provided in Appendix \ref{app:breakouts}.

As mentioned earlier, both the Prawnblaster and PrawnDO use a standard serial interface that is amenable for use in any control system.
However, we have designed both boards specifically for use in the \labscript experimental control suite.
To facilitate this, we have also provided interfacing code to use both devices within the \labscript experimental control system.
The timing for \labscript already relies on the concept of pseudoclocks, and our interfacing code automatically handles the details of triggering between boards, including accounting for triggering delays.

There are some obvious avenues for improvement of both firmwares.
First is to further improve write speeds of the pulse sequences, which are presently limited by the serial over USB interface of the RP2040.
While the bulk binary write speed is fast enough to write the entire available memory in approximately \qty{300}{\milli\second}, this can still be an issue for experiments with a similar duration.
In this case, programming large numbers of instructions could limit experiment cycling time.
Therefore, improvements to the underlying communication speeds would allow for even faster iteration.

For very large pulse sequences, leveraging the Pico's flash memory to increase the number of instructions could also be of benefit.
Reads from the flash memory are nearly as fast as from the SRAM, meaning performance would not be hindered.
However, storing instructions to the flash memory would be complicated as it has a minimum block erase size of \qty{4}{\kilo\byte} which dictates how many instructions must be bundled to allow clean reading and writing of data.

Finally, the RP2040 contains standardized functionality for I2C or SPI inter-communication between Pico boards, which could allow for a single interfacing board to program and control many sub-boards.
This would reduce the number of dedicated USB connections required in a complicated system of many boards
and would allow for a unified pulse specification for the combined system.

\section{Conclusion}

Timing in experimental control systems is a common and challenging problem.
Flexible, low-cost, scalable digital pulse generation systems to implement timing triggers for a larger system are desirable as experiments and systems increase in size and complexity.
While FPGAs provide the ultimate level of performance, flexibility, and functionality, they can be cost-prohibitive and often have more functionality than many experiments require.
Here we have shown how the commercially-available Raspberry Pi Pico board, which uses the RP2040 microcontroller unit, can be used to provide arbitrary digital pulse generation for experiment timing purposes.
These boards are widely available, low-cost, and can provide a level of performance that is sufficient for many experiments.
The Prawnblaster and PrawnDO firmwares we have developed will help expand the availability of high performance experimental timing and allow for significant scalability.

\begin{acknowledgments}
PM acknowledges support from the National Security Scholars Internship Program (NSSSIP).
CT acknowledges support from the National Science Foundation Graduate Research Fellowship Program (NSF GRFP) Grant No. 2141064.

The views, opinions and/or findings expressed are those of the authors and should not be interpreted as representing the official views or policies of the Department of Defense or the U.S. Government.
\end{acknowledgments}

\section*{Author Declarations}

\subsection*{Conflict of Interest}

The authors have no conflicts to disclose.

\subsection*{Author Contributions}

\textbf{P. T. Starkey}: Conceptualization (lead); Software (lead); Writing -- review and editing (equal).
\textbf{C. Turnbaugh}: Conceptualization (supporting); Software (supporting); Investigation (equal); Validation (equal); Writing -- review and editing (equal).
\textbf{P. Miller}: Software (supporting); Investigation (equal).
\textbf{K.-J. LeBlanc}: Investigation (equal); Validation (equal).
\textbf{D. H. Meyer}: Conceptualization (supporting); Software (supporting); Validation (equal); Visualization (lead); Writing -- original draft (lead); Writing -- review and editing (equal).

\section*{Data Availability Statement}

The data presented are available from the corresponding author upon reasonable request.

\appendix

\section{Breakout Board Design Considerations}
\label{app:breakouts}

Given the digital edges of the Prawnblaster and PrawnDO have transition times on the order of nanoseconds, special considerations should be taken when incorporating with other devices in an experimental control system.
In general, any digital pulse generation in an experimental control system will need to be compatible with 5V TTL, able to drive long \qty{50}{\ohm} impedance cables, and minimize cross-talk between outputs.
Though we do not intend to prescribe a particular solution to these challenges, we have designed custom breakout boards to mitigate these challenges during development and testing of the Prawnblaster and PrawnDO devices.
The designs are available online \cite{prawnblaster_breakout,prawndo_breakout}.

Our designs use surface-mount 74LVT and 74LVC series unity-gain buffers and line-drivers to provide the necessary compatibility with 5V TTL and drive currents for large capacitive loads.
All inputs and outputs are provided by SMA connections, with all outputs series terminated with \qty{50}{\ohm} and all inputs terminated into a 470/\qty{47}{\kilo\ohm} voltage divider.
Finally some care has been taken in the board layout to minimize cross-talk between outputs.

The data of Figure \ref{fig:measTraces} used boards fabricated in-house using a CNC mill, and therefore only have copper cladding on one side of the dielectric substrate.
This compromise results in longer ground return paths which has some impact on the cross-talk performance.

\section{Example Trace Programs}
\label{app:programs}

The pseudoclock program used in the blue trace of Figure \ref{fig:measTraces} is shown in Table \ref{tab:prawnblasterProgram}.
\begin{table}[h]
    \centering
    \begin{tabular}{cc}\toprule
        half period & repetitions \\
        \midrule
        90 & 3 \\
        5 & 20 \\
        100 & 1 \\
        10 & 3 \\
        50 & 2 \\
        0 & 0 \\
        \bottomrule
    \end{tabular}
    \caption{Pseudoclock program of the blue trace in Fig.~\ref{fig:measTraces}. Note that numbers are base-10 integers.}
    \label{tab:prawnblasterProgram}
\end{table}
Each line represents a successive instruction programmed into a single pseudoclock of the Prawnblaster.
The first number is the half-period of each tick of the pseudoclock as the number of clock cycles. The second number is the number of ticks to produce. These values are shown in base-10, which is the what the Prawnblaster expects.

The PrawnDO program used for the orange, green, and red traces is shown in Table \ref{tab:prawndoProgram}.
Each line represents a successive instruction programmed into the PrawnDO. 
\begin{table}[ht]
    \centering
    \begin{tabular}{cc}\toprule
       output state  & dwell time \\
       \midrule
    7 & 2D\\
    6 & 32\\
    5 & 32\\
    6 & 32\\
    5 & 32\\
    1 & 15E\\
    4 & 5\\
    6 & 6\\
    7 & 5\\
    6 & 7\\
    4 & 5\\
    3 & 7\\
    2 & 5\\
    4 & 5\\
    6 & 5\\
    5 & 5\\
    4 & 5\\
    7 & 5\\
    6 & 1E\\
    4 & 1E\\
    7 & F\\
    4 & A0\\
    6 & 64\\
    3 & 12C\\
    0 & 0\\
    0 & 0\\
    \bottomrule
    \end{tabular}
    \caption{PrawnDO program of the orange, green, and red traces of Fig.~\ref{fig:measTraces}. Note that numbers are base-16 hexidecimal.}
    \label{tab:prawndoProgram}
\end{table}
Here the first number is the output state of the 16 outputs of the PrawnDO, expressed as a base-16 hexidecimal.
The second number is the number of clock cycles to hold the output state, also expressed as a base-16 hexidecimal.
Note that the first instruction has been programmed to be five fewer clock cycles. This is to account for the trigger delay of the PrawnDO relative to the starting trigger pulse of the Prawnblaster.

\bibliography{Prawns}

\end{document}